\documentclass[12pt]{article}
\usepackage{graphicx}
\catcode`\@=11
\newcommand{\beq}{\begin{equation}}
\newcommand{\eeq}{\end{equation}}
\newcommand{\abs}[1]{\vert#1\vert}
\newcommand{\eps}{\varepsilon}
\renewcommand{\flat}{{\rm flat}}

\newcommand{\mean}[1]{\langle#1\rangle}
\newcommand{\prob}{\mathop{\rm Prob}}
\newcommand{\s}{\sigma}
\newcommand{\sign}{\mathop{\rm sign}}
\newcommand{\xidy}{\xi_{\rm dyn}}
\newcommand{\C}{{\cal C}}
\newcommand{\N}{{\cal N}}
\newcommand{\T}{\Gamma}
\begin{document}
\centerline{\bf A/161045/LET -- Revised}
\vspace{1.4cm}
\centerline{\Large\bf Why shape matters in granular compaction}
\vspace{1.6cm}
\centerline{\large
Anita Mehta$^{a,}$\footnote{anita@bose.res.in}
and J.M.~Luck$^{b,}$\footnote{luck@spht.saclay.cea.fr}}
\vspace{1cm}
\centerline{$^a$S.N.~Bose National Centre for Basic Sciences,}
\centerline{Block JD, Sector 3, Salt Lake, Calcutta 700098, India}
\centerline{$^b$Service de Physique Th\'eorique\footnote{URA 2306 of CNRS},
CEA Saclay, 91191 Gif-sur-Yvette cedex, France}
\vspace{1cm}
\begin{abstract}
We present a stochastic model
of dynamically interacting grains in one dimension,
in the presence of a low vibrational intensity,
to investigate the effect of shape
on the statics and dynamics of the compaction process.
Regularity and irregularity in grain shapes are shown
to be centrally important in determining the statics of close-packing states,
as well as the nature of zero- and low-temperature dynamics
in this columnar model.
\end{abstract}
\vfill
\noindent P.A.C.S.: 45.70.-n, 61.43.Fs, 64.60.Cn, 64.70.Pf, 75.10.Hk
\newpage

That shape matters crucially in the efficient packing
of objects is everyday experience.
Coffee grains can be shaken into a closely packed state
far more readily than marbles,
where necks and pores remain even in the most ordered state.
This Letter is devoted to a simple stochastic dynamical model
in one dimension,
demonstrating the effect of shape in closely packed systems,
such as glasses or densely packed granular media
(see~\cite{pgg} for a review, and~\cite{lattice}
for recent investigations of lattice models of granular systems).
A more general version of the model, where lattice sites can be empty,
has been described elsewhere~\cite{usepl,usproc}.
The present model, and some of its generalisations,
will be investigated in more detail in~\cite{long}.

\subsubsection*{The model}

Our model consists in a column of $N$ sites,
each of which is occupied by a grain.
Grains are indexed by their depth $n=1,\dots,N$,
measured from the top of the column.
Grains are assumed to be anisotropic in shape.
They take, for simplicity, two possible orientations,
referred to as ordered and disordered.
We define orientation variables
by setting $\s_n=+1$ if grain number $n$ is ordered,
and $\s_n=-1$ if grain number $n$ is disordered.
A configuration of the system is uniquely defined
by the $N$ orientation variables $\{\s_n\}$.
While ordered grains are perfectly packed,
disordered grains are imperfectly packed.
Each disordered grain leaves a void space $\eps$ on the site it inhabits.

Non-trivial dynamical interactions
between the grains are such as to minimise the void space locally.
These lead to frustration between
competing grain orientations which seek to minimise interstitial voids.
Under the influence of a dimensionless vibration intensity~$\T$,
this system undergoes cooperative reorganisations reminiscent of glassy
dynamics~\cite{glassyref}.
From a phenomenological viewpoint,
we model the above by a stochastic dynamics,
defined by the following transition rates per unit time:
\beq
\left\{\matrix{
w_n(\s_n=+\to\s_n=-)=\exp\left(-n/\xidy-h_n/\T\right),\hfill\cr\cr
w_n(\s_n=-\to\s_n=+)=\exp\left(-n/\xidy+h_n/\T\right).\hfill}\right.
\label{w}
\eeq
The dynamical rules~(\ref{w}) fully define the model.
In these formulas, $\T$ is a dimensionless measure of the vibration intensity,
which will also be referred to as temperature.
The dynamical length $\xidy$ is a phenomenological parameter
which controls the spatial dependence of dynamical behaviour~\cite{long}.
(In earlier work~\cite{usepl} $\xidy$ was shown to determine the extent to
which
order propagates down the column in the glassy regime, a perspective
that is maintained here).
Finally, and most importantly, the ordering field $h_n$
acting on the orientation $\s_n$ of grain number $n$ is {\it compacting}
(see later): it reads
\beq
h_n=\eps\,m^-_n-m^+_n,
\label{ydef}
\eeq
where $m^+_n$ (resp.~$m^-_n$)
is the number of ordered (resp.~disordered) grains above grain number~$n$:
\beq
m^+_n=\frac12\sum_{k=1}^{n-1}(1+\s_k),\qquad
m^-_n=\frac12\sum_{k=1}^{n-1}(1-\s_k),
\eeq
so that $m^+_n+m^-_n=n$.
Hence~$h_n$ represents
the excess void space~\cite{brownrichards} of the system.
Equation~(\ref{ydef}) shows that a transition
from an ordered to a disordered state for grain number~$n$ is
{\it hindered} by the number of voids that are already above it,
i.e., represents a compacting dynamics.
Our model is thus fully directed in time and space:
the orientation of a given grain only influences the grains below it
and at later times.
The free surface of such a system under vibration could be expected to have
maximal mobility;
the free volume generated by vibration~\cite{bm} would
decrease with increasing depth in the column.

In the limit of a vanishing vibration intensity ($\T\to0$),
the probabilistic rules~(\ref{w}) become the following deterministic
updating formula:
\beq
\s_n\to\sign\,h_n=\sign(\eps\,m^-_n-m^+_n),
\label{zerody}
\eeq
provided $h_n\ne0$~(see below).

\subsubsection*{Statics}

Ground states are defined as being the configurations
such that the static counterpart of~(\ref{zerody}):
\beq
\s_n=\sign\,h_n=\sign(\eps\,m^-_n-m^+_n),
\label{zerost}
\eeq
holds for all grains $n=1,\dots,N$.
In other words, in a ground state
each orientation $\s_n$ is aligned along its local field $h_n$.

We mention for completeness that the case $\eps<0$
is a generalisation of earlier work~\cite{usepl},
with a complete absence of frustration and a single
ground state of ordered grains.
In the present situation ($\eps>0$),
a rich ground-state structure is achieved,
because of frustration~\cite{sg},
whose nature depends on whether $\eps$ is rational or irrational.
The rotation number
\beq
\Omega=\eps/(\eps+1)
\eeq
fixes the proportions of ordered and disordered grains
in the ground states: $f_+=\Omega$, $f_-=1-\Omega$, and their difference
$\mean{\s_n}=f_+-f_-=2\Omega-1$ can be seen as a spatially
averaged `magnetisation'.

\medskip
\noindent $\bullet$
For irrational $\eps$,~(\ref{ydef}) implies that
all the local fields $h_n$ are non-zero.
A unique quasiperiodic ground state is thus generated~\cite{long}.
It can be constructed by the cut-and-project method,
along the lines of a geometrical approach
developed for quasicrystals~\cite{quasi}.
The local fields $h_n$ lie in a bounded interval $-1\le h_n\le\eps$.

\medskip
\noindent $\bullet$
For rational $\eps=p/q$, with $p$ and~$q$ mutual primes,
$\Omega=p/(p+q)$, and some of the $h_n$ can vanish.
The equation $h_n=0$ means that grain number~$n$
has a perfectly packed column above it,
so that it is free to choose its orientation.
For $\eps=1/2$, for example, one can visualise
that each disordered grain carries a void half its
size, so that units of perfect packing must be permutations
of the triad $+--$, where the two half voids from each
of the $-$ grains are filled by the $+$ grain.
The dynamics selects two of these patterns, $+--$ and $-+-$.
More generally, orientational indeterminacy occurs at points
of perfect packing such that~$n$ is a multiple of the period $p+q$.
Each ground state is a random sequence of two patterns of length $p+q$,
each containing $p$ ordered and~$q$ disordered grains.
The model therefore has a zero-temperature configurational entropy
(or ground-state entropy) per grain:
\beq
\Sigma=\ln 2/(p+q).
\eeq

Associating (ir)regular grains with (ir)rational void spaces,
this has the appealing physical interpretation that irregularities
in grain shapes lead to a unique state of close packing
(such that all jagged edges are well meshed together),
while regular grains have a huge degeneracy of such states
(as in the fabled greengrocer's problem~\cite{greengrocer}).

\subsubsection*{Zero-temperature dynamics}

We now turn to the zero-temperature dynamics~(\ref{zerody}),
starting with a disordered initial configuration.

\medskip
\noindent $\bullet$
For irrational $\eps$,
the application of this dynamics causes the
quasiperiodic ground state to be recovered
downwards from the top of the column.
The depth of the ordered boundary layer grows ballistically with time:
\beq
L(t)\approx V(\eps)\,t.
\eeq
The velocity $V(\eps)=V(1/\eps)$ varies smoothly with $\eps$,
and diverges as $V(\eps)\sim\eps$ for $\eps\gg1$~\cite{long}.
The rest of the system remains in its disordered initial state.
When $L(t)$ becomes comparable with $\xidy$,
the effects of the free surface begin to be damped.
In particular for $t\gg\xidy/V(\eps)$
we recover the logarithmic coarsening law $L(t)\approx\xidy\ln t$,
observed in related work~\cite{usepl,bergmehta}
to model the slow dynamical relaxation of vibrated sand~\cite{sid}.

\medskip
\noindent $\bullet$
For rational $\eps$, as mentioned above, the local field $h_n$ may vanish.
The corresponding orientation
is updated according to $\s_n\to\pm1$ with probability $1/2$, leading to a
dynamics which is stochastic even at zero temperature.
Here, even the behaviour well within the boundary layer
$\xidy$ contains many intriguing features, while the dynamics for $n\gg\xidy$
again takes place on a logarithmic scale~\cite{long}.
We therefore focus on the limit $\xidy=\infty$.
The main result is that zero-temperature dynamics
does not drive the system to any of its degenerate ground states.
The system instead shows a fast relaxation to a non-trivial steady state,
independent of its initial condition.
The local fields $h_n$ have unbounded fluctuations in this steady state,
which are reminiscent of the density fluctuations
about the mean packing fraction observed in granular systems~\cite{sid,gcbam},
especially around the so-called
random close packing density~\cite{brownrichards,bernal}, which
is the highest density achievable in practice by extensive dynamical processes.

\begin{figure}[htb]
\begin{center}
\includegraphics[angle=90,width=.6\linewidth]{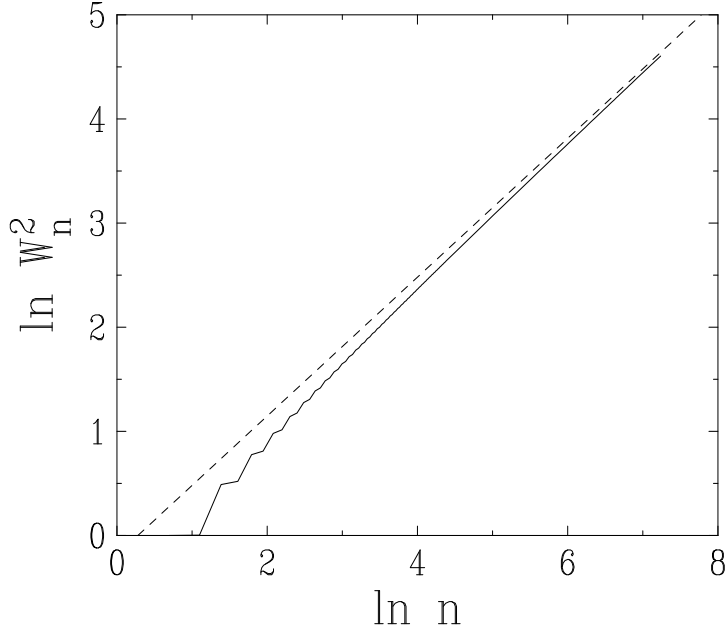}
\caption{\small
Log-log plot of $W_n^2=\mean{h_n^2}$ against depth $n$,
for zero-temperature dynamics with $\eps=1$.
Full line: numerical data.
Dashed line: fit of asymptotic behaviour, leading to~(\ref{rough}).}
\label{fign}
\end{center}
\end{figure}

Figure~\ref{fign} demonstrates the anomalous roughening law
\beq
W_n^2=\mean{h_n^2}\approx A\,n^{2/3}\qquad(A\approx 0.83).
\label{rough}
\eeq
The local fields $h_n$ are approximately Gaussianly distributed.
The following scaling argument explains the observed roughening exponent $2/3$.
Let $h_n$ be the position of a fictitious random walker at time $n$.
The noise in this random walk
originates in the sites $m<n$ where the local field~$h_m$ vanishes.
It is therefore proportional to $\sum_{m=1}^{n-1}\prob\{h_m=0\}$,
hence the consistency condition $W_n^2\sim\sum_{m=1}^{n-1}1/W_m$,
yielding the power law~(\ref{rough}).
A more detailed derivation will be given in~\cite{long}.
Roughly speaking, the walker obeys a diffusion equation,
with a diffusion constant scaling as the inverse distance to its starting
point.
This is reminiscent of the domain-growth mechanism
in the low-temperature coarsening regime of the Ising chain
with Kawasaki dynamics~\cite{cks}: the power law $L(t)\sim t^{1/3}$
for the mean domain size (analogous to $W_n\sim n^{1/3}$)
can be understood from the picture of diffusing domains,
whose diffusion constant scales as the inverse of their length.
The anomalous roughening law~(\ref{rough})
is the most central feature of the zero-temperature steady state
observed for rational $\eps$.

\begin{figure}[htb]
\begin{center}
\includegraphics[angle=90,width=.6\linewidth]{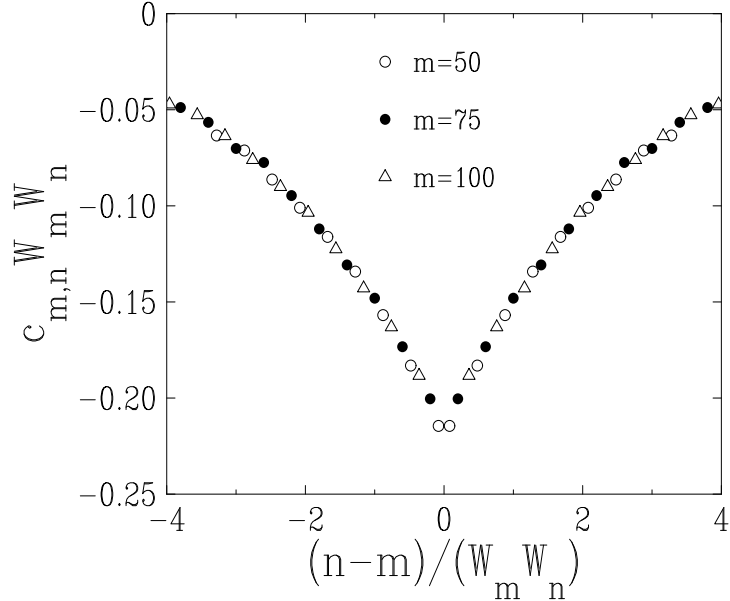}
\caption{\small
Scaling plot of the orientation correlation function $c_{m,n}$ for $n\ne m$
in the zero-temperature steady state with $\eps=1$,
demonstrating the validity of~(\ref{cmn}),
and showing a plot of (minus) the scaling function $F$.}
\label{figo}
\end{center}
\end{figure}

If the grain orientations were statistically independent, i.e., uncorrelated,
one would have the simple result $\mean{h_n^2}=(\eps^2+1)n/2$,
while~(\ref{rough})
implies that $\mean{h_n^2}$ grows much more slowly than $n$.
Orientational displacements are thus fully anticorrelated.
More precisely, Figure~\ref{figo} shows that
the orientation correlations $c_{m,n}=\mean{\s_m\s_n}$ scale as~\cite{long}
\beq
c_{m,n}\approx\delta_{m,n}
-\frac{1}{W_mW_n}\,F\!\left(\frac{n-m}{W_mW_n}\right).
\label{cmn}
\eeq
The scaling function $F$ is even, positive, and it obeys
$\int_{-\infty}^{+\infty}F(x)\,{\rm d}x=1$,
expressing that spin fluctuations are asymptotically totally screened
(i.e., fully anticorrelated).
This is similar to the bridge collapse seen in displacement-displacement
correlations of strongly compacting grains~\cite{gcbam};
grain orientational displacements in the direction of vibration
were there seen to be strongly anticorrelated
in jammed regions, as each grain tried to collapse
into the void space trapped by its neighbours.
We remark that temporal anticorrelations have also been observed in
recent experiments investigating the properties of cages
near the colloidal glass transition~\cite{weeks}.
Interestingly, correlations transverse to the shaking direction
were found to be rather small~\cite{gcbam}, thus, in
self-consistency terms justifying the choice of a column
model in the present case.

\begin{figure}[htb]
\begin{center}
\includegraphics[angle=90,width=.6\linewidth]{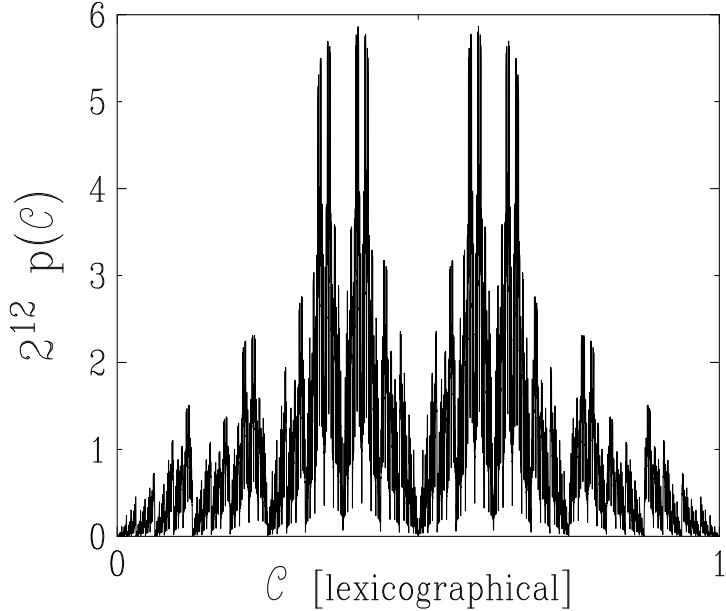}
\caption{\small
Plot of the normalised probabilities $2^N\,p(\C)$
of all the configurations in the zero-temperature steady state
for $N=12$ and $\eps=1$,
against the configurations $\C$ sorted lexicographically.}
\label{figi}
\end{center}
\end{figure}

Figure~\ref{figi} shows a plot of the normalised probabilities $2^N\,p(\C)$
of all the configurations~$\C$ in the zero-temperature steady state
of a system with $N=12$ grains and $\eps=1$,
directly measured in a very long simulation.
The probabilities are plotted against the $2^{12}=4096$ configurations,
sorted lexicographically (read down the column).
This plot exhibits a rugged structure on this microscopic scale:
some configurations are clearly visited far more often than others.
We suggest that this behaviour is generic: i.e.,
the dynamics of compaction in the jammed state leads to a microscopic
sampling of configuration space which is highly non-uniform.
In spite of this fine structure, the steady-state entropy
\beq
S=-\sum_{\C}p(\C)\ln p(\C),
\label{sdyn}
\eeq
is not far from the estimate $S_\flat=N\ln 2$
along the lines of Edwards' flatness hypothesis~\cite{sam}.
Indeed, for $N=12$ (data of Figure~\ref{figi}),
we have $S=7.839$ against $S_\flat=8.318$.
In other words, the entropy reduction~\cite{remi}
$\Delta S=S_\flat-S=0.479$ is small,
compared to the absolute value of the entropy per grain.

\subsubsection*{Low-temperature dynamics}

Low-temperature dynamics induces drastic changes in the case
of $\eps$ irrational.
(Since zero-temperature dynamics remains stochastic
in the rational case, we do not expect low-temperature
dynamics to introduce qualitative differences there.)
For a low but non-zero~$\T$, there will be a few mistakes,
i.e., orientations which are not aligned with their local field
according to~(\ref{zerost}).
The a priori probability of observing a mistake at site $n$ scales as
\beq
\Pi(n)\sim\exp\left(-2\abs{h_n}/\T\right).
\eeq
Hence the sites $n$ so that $\abs{h_n}\sim\T\ll1$
will be preferred nucleation sites for mistakes,
and thus dominate the low-temperature dynamics.
It turns out that those sites are such that $n\Omega$ is close to an integer.
The magnitude of the excess void space is least here,
leading to the least cost for a misalignment.

Choosing the golden mean $\eps=\Phi=(\sqrt{5}+1)/2$ for concreteness,
the preferred nucleation sites are given by the Fibonacci numbers:
$n=F_k\approx\Phi^k/\sqrt{5}$.
Then $\Pi(F_k)\sim\exp(-2\Phi^2/(\sqrt{5}\,\T F_k))$.
Let us denote the instantaneous position of the uppermost mistake by $\N(t)$.
This depth divides an upper boundary layer,
ordered in the quasiperiodic ground state, from a disordered lower zone,
characterised by the anomalous roughening law~(\ref{rough}).
The mistake is itself advected ballistically with velocity
$V(\Phi)\approx2.57$, just as with zero-temperature dynamics,
until another mistake is nucleated above it.

\begin{figure}[htb]
\begin{center}
\includegraphics[angle=90,width=.6\linewidth]{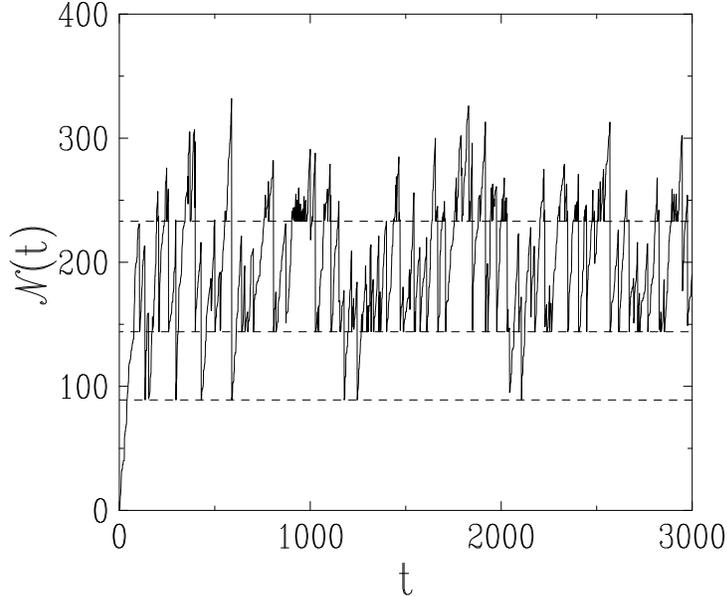}
\caption{\small
Plot of the instantaneous depth $\N(t)$ of the ordered layer,
for $\eps=\Phi$ (the golden mean) and $\T=0.003$.
Dashed lines: leading nucleation sites given by Fibonacci numbers
(bottom to top: $F_{11}=89$, $F_{12}=144$, $F_{13}=233$).}
\label{figk}
\end{center}
\end{figure}

Figure~\ref{figk} shows a typical sawtooth plot of the instantaneous
depth $\N(t)$, for a temperature $\T=0.003$.
The ordering length $\mean{\N}$ is expected to diverge at low temperature,
as mistakes become more and more rare.
From a more quantitative viewpoint, the most active Fibonacci site $F_k$
is such that the nucleation time $1/\Pi(F_k)$
is comparable to the advection time to the next nucleation site $F_{k+1}$.
This yields
\beq
\mean{\N}\sim1/(\T\,\abs{\ln\T}).
\eeq
The ordering length therefore diverges linearly at low temperature,
up to a logarithmic correction.
A similar law is predicted for all the irrational values of $\eps$
with typical Diophantine properties~\cite{long}.
This mechanism, where mistakes appear at preferred nucleation sites,
is reminiscent of the phenomenon of hierarchical melting
observed in incommensurate modulated solids~\cite{hime}.

Both $\mean{\N}$ and $\xidy$ retain the flavour of a boundary layer separating
order from disorder.
Within each of these boundary
layers, the relaxation is fast, and based on
single-particle relaxation, i.e., individual particles
attaining their positions of optimal local packing~\cite{bergmehta,gcbam}.
The slow dynamics of
cooperative relaxation only sets in for lengths beyond these,
when the lengths over which packing needs to be optimised become non-local.
This in turn leads, as in reality~\cite{brownrichards}, to hysteresis,
i.e., a dependence on the initial state of the packing.

Remarkably, all of these features were obtained at a qualitative
level in the glassy regime of a much simpler model~\cite{usepl}.
On the one hand, this allows us to speculate
that the shape-dependent aging phenomena
seen there could be retrieved here, i.e.,
that conventional aging phenomena would
only be seen for irregular grains (irrational $\eps$).
On the other hand, it is tempting to ask if the directional causality
of the dynamical interactions
present in this model and the earlier one~\cite{usepl}
could be responsible for their qualitative similarity, and thus
be a necessary ingredient for modelling glassiness.


\begin{thebibliography}{99}

\bibitem{pgg}
P.G. de Gennes, Rev. Mod. Phys. {\bf 71}, S374 (1999).

\bibitem{lattice}
A. Fierro, M. Nicodemi, and A. Coniglio, Phys. Rev. E {\bf 66}, 061301 (2002);
R.C. Ball and R. Blumenfeld, Phys. Rev. Lett. {\bf 88}, 115505 (2002);
I. Goldhirsch and C. Goldenberg, Eur. Phys. J. E {\bf 9}, 245 (2002).

\bibitem{usepl}
P.F. Stadler, J.M. Luck, and A. Mehta, Europhys. Lett. {\bf 57}, 46 (2002).

\bibitem{usproc}
P.F. Stadler, A. Mehta, and J.M. Luck,
Adv. Complex Systems {\bf 4}, 429 (2001).

\bibitem{long}
J.M. Luck and A. Mehta (in preparation).

\bibitem{glassyref}
M.F. Shlesinger and J.T. Bendler, in {\it Phase Transitions in Soft Condensed
Matter}, T. Riste and D. Sherrington, eds. (Plenum, 1989);
R. Monasson, Phys. Rev. Lett. {\bf 75}, 2847 (1995);
E. Marinari, G. Parisi, F. Ricci-Tersenghi, and F. Zuliani,
J. Phys. A {\bf 34}, 383 (2001);
L. Berthier, L.F. Cugliandolo, and J.L. Iguain,
Phys. Rev. E {\bf 63}, 051302 (2001);
M. M\'ezard, Physica A {\bf 306}, 25 (2002);
G. Biroli and M. M\'ezard, Phys. Rev. Lett. {\bf 88}, 025501 (2002);
A. Lawlor, D. Reagan, G.D. McCullagh, P. De Gregorio, P. Tartaglia,
and K.A. Dawson, Phys. Rev. Lett. {\bf 89}, 245503 (2002).

\bibitem{brownrichards}
R.L. Brown and J.C. Richards, {\it Principles of Powder Mechanics}
(Pergamon, Oxford, 1970).

\bibitem{bm}
A. Mehta and G.C. Barker, Phys. Rev. Lett. {\bf 67}, 394 (1991);
Europhys. Lett. {\bf 27}, 501 (1994).

\bibitem{sg}
For a review, see: M. M\'ezard, G. Parisi, and M.A. Virasoro,
{\it Spin Glass Theory and Beyond} (World Scientific, Singapore, 1987).

\bibitem{quasi}
N.G. de Bruijn, Nederl. Akad. Wetens. Proc. A {\bf 84}, 27 (1981);
M. Duneau and A. Katz, Phys. Rev. Lett. {\bf 54}, 2688 (1985);
J. Phys. (France) {\bf 47}, 181 (1986);
V. Elser, Phys. Rev. B {\bf 32}, 4892 (1985);
P.A. Kalugin, A.Yu. Kitayev, and L.S. Levitov, JETP Lett. {\bf 41}, 145 (1985);
J. Phys. (France) Lett. {\bf 46}, L601 (1985).

\bibitem{greengrocer}
S. Torquato, {\it Random Heterogeneous Materials:
Micro\-struc\-ture and Macroscopic Properties} (Springer, New York, 2001).

\bibitem{bergmehta}
J. Berg and A. Mehta, Europhys. Lett. {\bf 56}, 784 (2001);
Phys. Rev. E {\bf 65}, 031305 (2002).

\bibitem{sid}
E.R. Nowak, J.B. Knight, E. Ben-Naim, H.M. Jaeger, and S.R. Nagel,
Phys. Rev. E {\bf 57}, 1971 (1998);
E.R. Nowak, A. Grushin, A.C.B. Barnum, and M.B. Weissman,
Phys. Rev. E {\bf 63}, 020301 (2001).

\bibitem{gcbam}
G.C. Barker and A. Mehta, Phys. Rev. A {\bf 45}, 3435 (1992);
Phys. Rev. E {\bf 47}, 184 (1993);
A. Mehta and G.C. Barker, J. Phys. Cond. Matt. {\bf 12}, 6619 (2000).

\bibitem{bernal}
J.D. Bernal, Proc. R. Soc. London A {\bf 280}, 299 (1964).

\bibitem{cks}
S.J. Cornell, K. Kaski, and R.B. Stinchcombe, Phys. Rev. B {\bf 44}, 12263
(1991);
S.J. Cornell and A.J. Bray, Phys. Rev. E {\bf 54}, 1153 (1996);
V. Spirin, P.L. Krapivsky, and S. Redner, Phys. Rev. E {\bf 60}, 2670 (1999).

\bibitem{weeks}
E.R. Weeks and D.A. Weitz, Chem. Phys. {\bf 284}, 361 (2002).

\bibitem{sam}
S.F. Edwards, in {\it Granular Matter: An Interdisciplinary Approach},
A. Mehta, ed. (Springer, New York, 1994).

\bibitem{remi}
R. Monasson and O. Pouliquen, Physica A {\bf 236}, 395 (1997).

\bibitem{hime}
F. Vallet, R. Schilling, and S. Aubry, Europhys. Lett. {\bf 2}, 815 (1986);
R. Schilling and S. Aubry, J. Phys. C {\bf 20}, 4881 (1987);
F. Vallet, R. Schilling, and S. Aubry, J. Phys. C {\bf 21}, 67 (1988).

\end{thebibliography}
\end{document}